\documentclass[letterpaper]{article} 
\usepackage{aaai25}  
\usepackage{times}  
\usepackage{helvet}  
\usepackage{courier}  
\usepackage[hyphens]{url}  
\usepackage{graphicx} 
\usepackage{natbib}  
\usepackage{caption} 
\usepackage{algorithm}
\usepackage{algorithmic}

\usepackage{newfloat}
\usepackage{listings}
\DeclareCaptionStyle{ruled}{labelfont=normalfont,labelsep=colon,strut=off} 
\floatstyle{ruled}
\newfloat{listing}{tb}{lst}{}
\floatname{listing}{Listing}
\title{The Power of Social Norms: How Initial Responses to Toxicity Shape Conversations on Twitter}

\author{
    Ana Aleksandric\textsuperscript{\rm 1}\thanks{Work done at The University of Texas at Arlington.}
    Mohit Singhal\textsuperscript{\rm 2}\footnotemark[1], Anne Groggel\textsuperscript{\rm 3}, 
    Shirin Nilizadeh\textsuperscript{\rm 4}
}
\affiliations{
    \textsuperscript{\rm 1}Florida Atlantic University\\
    \textsuperscript{\rm 2}Northeastern University\\
    \textsuperscript{\rm 3}St. Mary's College of Maryland\\
    \textsuperscript{\rm 4}The University of Texas at Arlington\\
aaleksandric@fau.edu, m.singhal@northeastern.edu, agroggel@smcm.edu, shirin.nilizadeh@uta.edu

}

\usepackage{caption}
\usepackage{subcaption}

\usepackage{xcolor}
\usepackage{bibentry}

\begin{document}
\maketitle
\begin{abstract}
Online harassment and abusive language continue to be a growing concern on social media platforms. In this study, we explore the power of group dynamics to shape the toxicity of Twitter conversations. 
First, we examine how the presence of others in a conversation can potentially diffuse Twitter users' responsibility to address a toxic reply. Second, we examine whether the toxicity of the first direct reply to a toxic tweet in conversations establishes group norms for subsequent replies. By doing so, we outline users participating in the conversation before the first toxic reply and the tone of initial responses to a toxic reply as explanatory factors that affect whether others feel uninhibited to post their own abusive or derogatory replies.
We test this premise by analyzing a random sample of more than 187K tweets belonging to $\sim 9K$ conversations. 
This analysis of group dynamics is motivated by a larger body of scholarship on contagion of antisocial behavior and the power of establishing social norms that maintain rather than sanction toxicity. 
We find evidence that an increased number of users participating in the conversation before receiving a toxic tweet is negatively associated with the number of users who responded to the toxic reply in a non-toxic way.
Furthermore, posting a toxic reply immediately after a toxic comment is negatively associated with users posting non-toxic replies and Twitter conversations becoming increasingly toxic. We argue that understanding how social media users respond to uncivil comments or abusive language reveals social norms as powerful social cues that can shape human behavior online.
\end{abstract}

\begin{center}
\large
\textcolor{red}{\textbf{This paper has been accepted at CySoc 2026, please cite accordingly.}
}\end{center}

\section{Introduction}
\vspace{1pt}
\noindent\fbox{%
    \parbox{.98\columnwidth}{%
        \textbf{Content Warning}: \textcolor{red}{This study analyzes group dynamics in online toxic conversations. This paper provides demonstrative examples of user content
        that might include profane and hateful content that may be found offensive by some.}
    }%
}
\vspace{2pt}

Social media and online communities allow individuals to freely express opinions, engage in interpersonal communication, and learn about new trends and news stories. Platforms such as Twitter (now known as X) hold promise for users to engage in rich and vibrant conversations with others from various backgrounds and cultures. Nevertheless, these platforms also serve as spaces for uncivil behavior.
In particular, toxicity as explicit language, derogatory, aggressive, or disrespectful content has become endemic on online platforms~\cite{anderson2016toxic,dutton1996network,papacharissi2002virtual,hwang2008does,zannettou2020measuring}. 
There is a growing concern regarding the prevalence of incivility over social media platforms and its impact on online communities~\cite{rost2016digital,duggan2014online}. How groups react to divisive behavior can reflect broader social norms online, whether they spread negative behaviors, call out racist or sexist behavior, or ignore toxic behaviors~\cite{binns2012don}. 

Toxic behavior frequently occurs in the presence of other users 
whose actions can influence the dynamics of a social situation. For instance, they may actively engage with the perpetrator's behavior by posting toxic replies, endeavor to counteract toxicity by confronting such behavior or contributing positively to the conversation, or simply observe the interaction unfold~\cite{aleksandric2024users}. 
Social norm theory provides a useful lens for understanding communication behavior on social media, particularly the persistence of toxicity. Norms reflect shared expectations about appropriate behavior, shaping the cultural “dos and don’ts” that guide interactions and regulate how individuals engage with content and other users online. In online communities, users are going to be guided both by their internalized discomfort with breaking social norms, their perceptions of what others view as acceptable, and by the behaviors they see others engaging in.

The belief that people behave differently in groups is a well-established social psychological tenet. 
In particular, the \emph{bystander effect theory} in social sciences refers to the phenomenon where individuals are less likely to offer help or intervene in emergency situations when other people are present~\cite{latane1969bystander,darley1968bystander}. This theory suggests that the presence of others can lead to diffusion of responsibility, where individuals believe that someone else will take action, resulting in a reduced likelihood of any single individual taking action themselves. The bystander effect has been studied extensively in psychology and sociology, and it highlights the complex social dynamics that influence human behavior in group settings~\cite{latane1968group,fischer2011bystander}.

However, less attention has been paid to how these dynamics play out in social media conversations, especially when a user becomes a target of toxic behaviors like harassment, hate speech, cyberbullying, or trolling. 
Previous literature has examined users' behavior in toxic conversations on social media~\cite{aleksandric2024users,xia2020exploring,shen2020viral}, highlighting that users tend to engage more in toxic than in non-toxic conversations. Also, Saveski et al.~\cite{saveski2021structure} explored how the likelihood of a toxic reply differs depending on whether the parent post is toxic
or not. Nevertheless, our understanding how the presence of others affects the user's behavior and their preference to encourage toxicity or stand up for the target remains underexamined.

Based on the bystander effect theory, the presence of others can diffuse one's sense of responsibility to help, with users believing another individual will act~\cite{latane1969bystander,darley1968bystander}.
Using social norms as a theoretical framework, we analyze a random sample of $\sim 9K$  \emph{Twitter conversations} 
to explore how group dynamics can influence online behavior.
We investigate how the presence of others in a conversation affects users' inclination to address toxic replies and how initial responses to such toxicity impact the overall tone of the conversation. To answer these questions, we conducted statistical tests while accounting for potential confounding factors, including users' account attributes, conversation structure, and topic of discussion. 

Our findings suggest that there is a negative, statistically significant relationship between the number of conversation participants before the first toxic reply and the number of unique users who respond to a toxic reply in a non-toxic way. This indicates that the greater the pre-toxic participation, the fewer users tend to engage in a non-toxic way after the first toxic reply occurs. In addition, the results demonstrate that the toxicity levels of the initial responses to toxic replies tend to affect the tone of the remainder of the conversation by establishing a norm. Moreover, qualitative analysis was employed to investigate how often positive standing up, i.e., correcting misunderstandings, agreeing civilly, or defending an individual or a group of people, occurs in toxic conversations. Through qualitative analysis, we find that users rarely attempted to resolve the conflict, with only 19.7\% standing up to resolve or stop the toxicity. 
Therefore, our findings suggest that group dynamics play an important role in shaping toxic threads, while some users' characteristics and discussion topics were also found relevant. 

In summary, this study sheds light on how group dynamics influence online behavior, which could be an initial step for developing effective interventions against toxicity.  In more detail, investigating how the presence of other users before the first toxic reply, as well as initial responses to toxicity, shape Twitter conversations, is crucial to this understanding. 
This comprehension can serve as a foundation for designing targeted interventions that leverage group dynamics to encourage positive engagement and mitigate the spread of harmful content. This contribution may help researchers understand how quickly antisocial norms can be established online and inspire future work to further investigate what factors lead users to adhere to or break social norms in addressing a toxic reply in civil conversations.

\section{Related Work}
\textbf{Social Norms: } 
Social norms are often conceptualized as prosocial, guiding interactions that benefit others or the collective while discouraging or sanctioning actions that cause harm~\cite{heckathorn1988collective}. Foundational work has found that some individuals adhere to group social norms even when these perceptions conflict with their own~\cite{asch1956studies}. 
But rather than promoting prosocial behavior, group dynamics can, in certain contexts, facilitate the spread of norm violations~\cite{alvarez2018normative}. Observing antisocial behaviors, such as littering or jaywalking, can lead individuals to perceive as accepted norms increasing the likelihood that they will engage in similar behaviors themselves~\cite{cialdini1991focus, mullen1990jaywalking}.

Research on cyberbullying has demonstrated that increasing the number of bystanders decreases intentions to intervene~\cite{obermaier2016bystanding}. However, other scholarship has shown that when individuals are aware they are visible to others, by using a webcam or making participants' screen-names more salient can reverse this effect~\cite{van2012aware}. 

Understanding the victim's perspective or empathizing with the target influence one's intentions of helping the victim ~\cite{paterson2019short,freis2013facebook,dominguez2018systematic}. 
In online communities, the adoption of antisocial behaviors may be amplified by the anonymity and reduced social cues these settings provide~\cite{suler2004online,lee2015people}, making individuals more likely to conform to group norms, even when those norms encourage harmful or antisocial conduct.

\textbf{Detection and Classification}. 
Empirical work on toxicity has employed machine learning-based detection algorithms to identify and classify offensive language, hate speech, and cyberbully~\cite{zhang2016conversational,davidson2017automated,koratanatoxic,pitsilis2018detecting,yin2023annobert,frenda2019online}. Works have used various methods ranging from lexical-based approaches~\cite{markov2021exploring,wiegand2018inducing} to deep neural networks~\cite{mazari2023bert,roy2021leveraging,del2023socialhaterbert,alshamrani2021hate,chen2019use,ribeiro2019inf}
Some recent works have used text and images together~\cite{yang2019exploring,singh2017toward} as well as text and socio-cultural information~\cite{vijayaraghavan2021interpretable} to detect hate speech. The state-of-the-art toxicity detection tool is available through Google's Perspective API~\cite{jigsaw2018perspective}. 
Perspective API has been studied and used extensively in the previous literature~\cite{kumarswamy2025causal,singhal2023sok,salehabadi2022,zannettou2020measuring,grondahl2018all,elsherief2018hate,saveski2021structure,kumar2023understanding,gonzalez2023understanding}. Hence, we will use Google Perspective API to detect toxic tweets in conversations.

\section{Hypotheses}

Prior work has examined the contagion of online toxicity in a group setting, such as focusing on team-based predictors of toxicity~\cite{shen2020viral}. Our work adds to the literature by investigating the group dynamics in naturally formed Twitter conversations, without predefined team structures or performance-based roles. Within our study, in larger conversations where no one speaks out against toxic behavior, participants may infer an implicit acceptance of such exchanges, reinforcing the perception that hostile replies are socially acceptable.
Therefore, we hypothesize: 

\textbf{H1}: The number of users participating in a conversation before the first toxic reply is negatively associated with the number of users who post non-toxic replies after the first toxic reply. 

We used the number of users who participated in the conversation before the first toxic reply as a proxy for users who are observing the conversation. The goal is to understand how the presence of a larger group shapes the remaining thread after the first toxic reply occurs. Typically, Twitter sends notifications to users when other accounts respond to their comments~\cite{twitterhelp-not}, so users can keep observing what happens in the thread.
Note that Twitter V2 provides the number of views a tweet has received. However, we did not use this metric because it is not a unique count~\cite{twview} and counts multiple views of the same user. 
Thus, the number of followers of the root authors was used as a control variable, since they might also observe the conversation. 

\textbf{Reactions to toxic behaviors:} Users observing the conversation can shape the reactions to toxic behavior online, whether they actively follow the perpetrator's behavior by posting additional toxic replies or attempting to re-establish norms of decorum. 
Online settings provide researchers with an opportunity to investigate how quickly norms are established among group members and the degree to which one's actions are expected to align with what is appropriate or expected within the group~\cite{tajfel2010social}. 
Once established, it is not surprising that individuals generally conform to group norms~\cite{asch1956studies}. 
Other work shows that observing trolling behavior by others influences new users~\cite{cheng2017anyone}. In other words, individuals may be more likely to post toxic replies after seeing others do so, believing it is the norm.

Peer conformity is positively associated with in-person bullying~\cite{duffy2009peer} and cyberbullying~\cite{ bleize2021social, bastiaensens2016normative}. 
In this instance, we might expect the toxicity of tweets within the conversation to alter based on the introduction and reaffirmation of toxic content. For example, passive behavior by users who initially participated in the conversation may be perceived as implicit approval of hate speech, while individual users' reactions, such as countering, are important in addressing online toxicity. 
For instance, when the first toxic tweet is met with negative sanctions, it may be perceived as inappropriate. In contrast, when initial toxicity is followed by further toxic replies, others may perceive such behavior as normative and follow suit. 
Thus, we test the following two hypotheses: 
\textbf{H2}: If a Twitter user posts a non-toxic reply immediately after the toxic reply, then more users post non-toxic replies. 
\textbf{H3}: If a Twitter user posts a non-toxic reply immediately after the toxic reply, then the toxicity of the conversation after this reply is more likely to be non-toxic.

These hypotheses test the premise that passive behaviors of users already participating in the conversation can be perceived as implicit approval of toxicity. Even though prior work found that the language toxicity of a comment significantly increased the number of its replies~\cite{xia2020exploring}, our analysis aims to shed light on whether the toxicity level of the initial response to the first toxic reply plays a significant role in toxicity levels of the remaining parts of the conversation.

\section{Data Collection}
\begin{figure*}[t]
    \centerline{\includegraphics[width=0.85\textwidth]{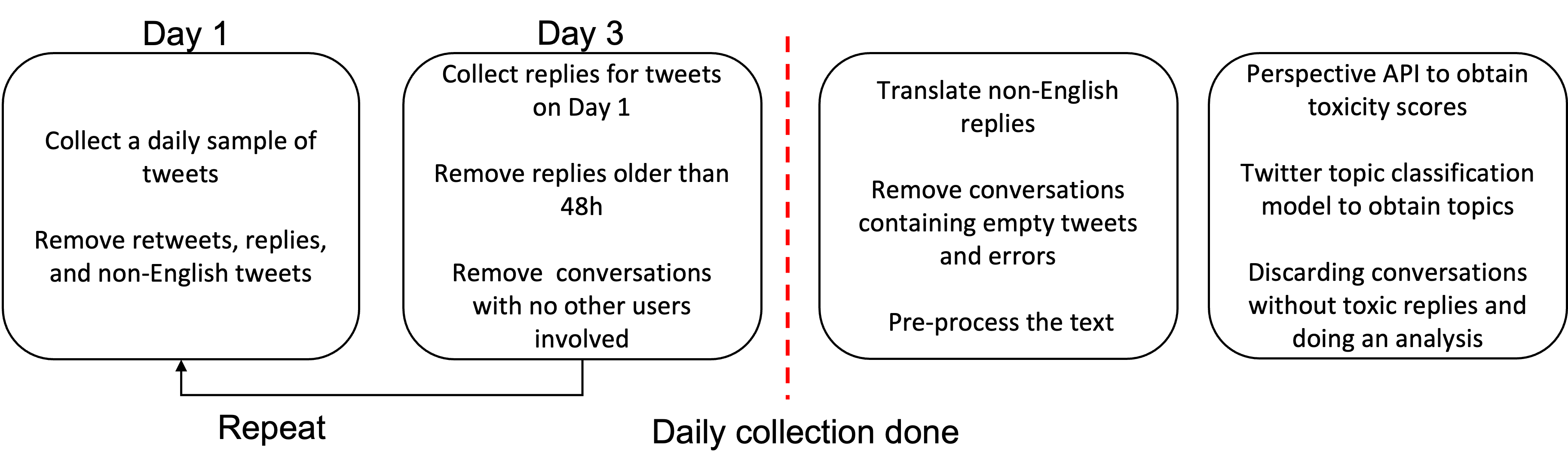}}
    \caption{Data collection flow. The figure demonstrates the timeline used for a collection of a random sample of tweets and their replies, as well as the data pre-processing steps.}
    \label{fig:collection}
\end{figure*}
\textbf{Dataset and Cleaning.}
Our dataset comes from Twitter\footnote{The data was collected before Twitter was rebranded to X, while Twitter API V2 was available. }, one of the most widely used social media platforms. One part of our data originates from a recent study~\cite{aleksandric2024users} where the dataset consists of a random sample of public English tweets during the period of August 14th to September 28th, 2021 (\emph{Dataset 1}). To expand our dataset, we used Twitter API V2~\cite{twitterapi} to gather an additional random sample of posts from March 31st to April 23rd, 2023 (\emph{Dataset 2}). A detailed data collection procedure for Dataset 2 is shown in Figure~\ref{fig:collection}. We collected two datasets to investigate if our findings can be generalized over a period of time. 

As described in Figure~\ref{fig:collection}, two days after the collection of daily tweets, we would use Twarc~\cite{twarc} to collect complete conversations for each initial tweet in the daily random sample. Twarc is a command-line tool and Python library for collecting Twitter data. 
We removed English tweets that are retweets or replies in other public conversations to only obtain the full conversations.
We also dismissed conversations in which all responses were posted by the author of the initial tweet, as no other users were involved. Moreover, the process of gathering conversation replies can take up to 24 hours, so some posts would have more time to acquire replies compared to others. 
Hence, we only saved replies received in the first 48 hours after the initial tweet was published, so the time frame for captured replies is consistent for all the conversations to reduce possible bias in the results. We used a threshold of 48 hours, as most of the replies occur closely after the initial tweet has been posted. 
We also removed conversations containing tweets or replies that only included links, images, or videos (named \emph{empty tweets}) rather than text, as Perspective API is a text-based toxicity detection tool~\cite{jigsaw2018perspective}. Furthermore, in certain cases, we were unable to extract the full conversation due to users hiding some of the replies or removing them. Such conversations were discarded from the dataset, as we would not certainly know whether the missing replies were toxic or who they belonged to. Finally, the cleaned dataset consists of 136,847 conversations containing 938,317 tweets.

\begin{figure}[t]
    \centerline{\includegraphics[width=0.7\columnwidth]{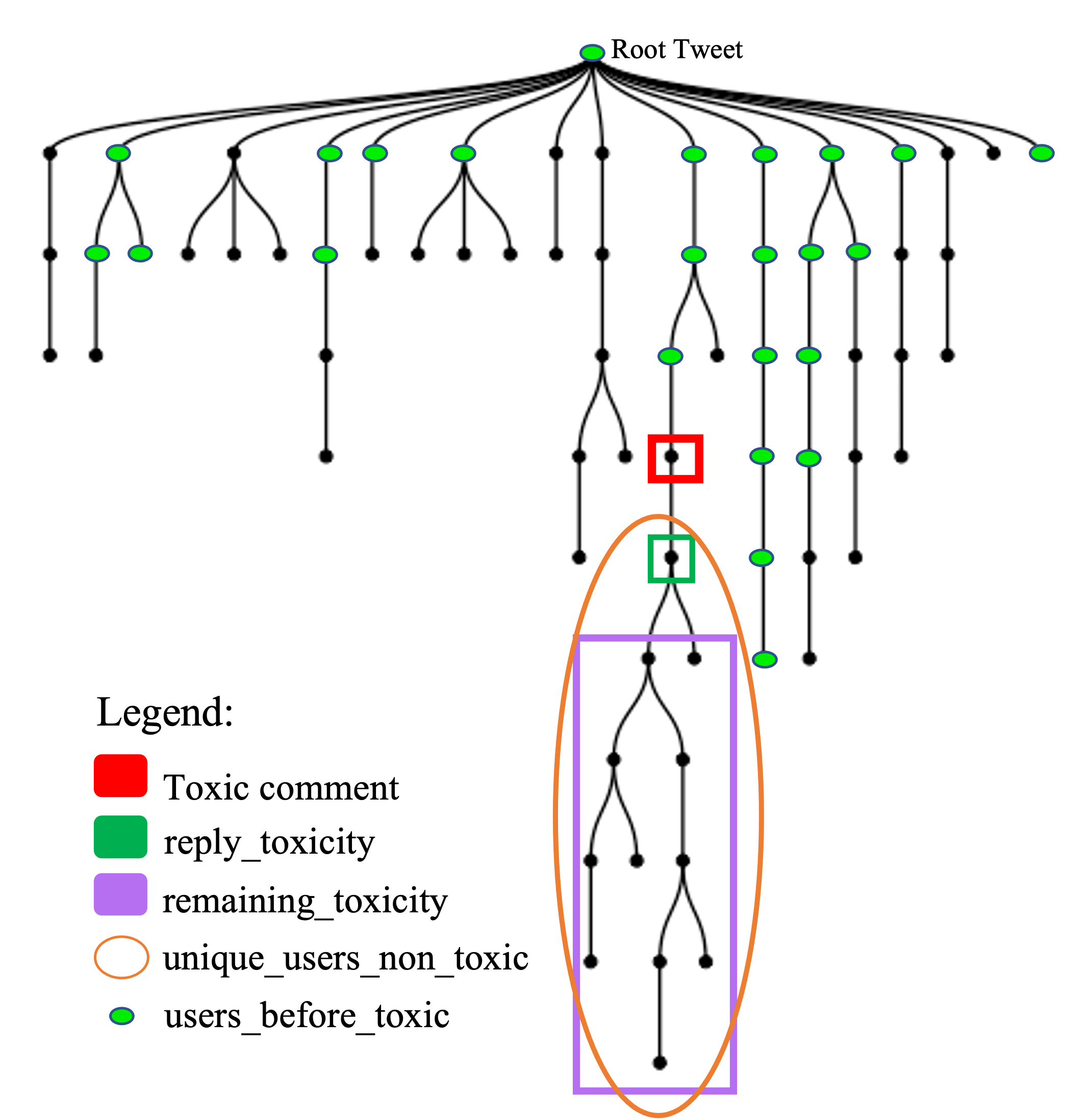}}
    \caption{An example of a conversation tree.}
    \label{fig:twittergraph}
\end{figure}

\textbf{Definitions.}
The user who posted the initial tweet is named \emph{root author}, and each full conversation is represented as a tree that started with the initial tweet named the \emph{root tweet}. As illustrated in Figure~\ref{fig:twittergraph}, each conversation is represented as a tree structure whose tweets (nodes) are connected when one is responding to another. 

\textbf{Obtaining Conversation Topic.} 
Certain topics, such as political ones, can trigger greater toxicity or attract more attention from other users, potentially revealing that the bystander effect tends to occur more frequently in specific conflicts. 
Thus, we used the Twitter multi-topic classification model trained on TweetTopic dataset, which is shared and evaluated by the recent study~\cite{antypas-etal-2022-twitter}, to obtain coefficients for 19 relevant discussion points on social media. Some topic examples include \emph{news \& social concern}, \emph{science \& technology}, \emph{diaries \& daily life}, \emph{business \& entrepreneurs}, etc. 
This RoBERTa-based model is fine-tuned on the Twitter corpus and has been recently used by several works~\cite{leiter2023chatgpt,hewitt2023backpack,cho2023deep,towle2023model}. 
We passed our initial tweets to the model as input to obtain scores in the range 0-1, where higher scores indicate a higher probability that the tweet is linked to a specific topic. 

\textbf{Discovering Conversations with Toxic Replies.} 
To identify all the toxic tweets in our dataset, we leveraged Google's Perspective API~\cite{jigsaw2018perspective}. Google Perspective API applies different machine learning models to score the toxicity of textual data. 
When a comment or text is passed as input to the Perspective API, the API returns scores for the requested attributes representing probability scores between 0 and 1. However, for this study, it is crucial to understand what such scores indicate. For example, a toxicity score of 0.8 means that 8 out of 10 people reading the comment would perceive the comment as toxic. To accomplish that, Google Perspective API is trained on millions of comments originating from multiple relevant sources, such as Wikipedia and The New York Times, across a range of languages~\cite{jigsaw2018perspective}. These comments are annotated by 3-10 coders who speak the suitable language, followed by using their labels to train the API models~\cite{jigsaw2018perspective}. Note that the models have been evaluated under the ROC curve and also checked for unintended biases for each of the identity groups~\cite{jigsaw2023perspective}.

In this paper, we consider scores for the \emph{Toxicity} attribute since Google's Perspective API defines a text having this attribute as a rude, disrespectful, or unreasonable comment that is likely to make people leave a discussion~\cite{jigsaw2018perspective}. 
Note that before passing our tweets to the Perspective API, we cleaned the text of tweets by removing punctuation and URLs, and replacing emojis with appropriate text describing the emojis. Moreover, the fact that our initial tweets are written in English does not guarantee that their replies will also be in English. Thus, we translated non-English replies into English, using Google Trans API~\cite{gggooo}. There were around 287K replies that were not classified as English replies by Twitter. 
Finally, we created a binary variable indicating whether a tweet was toxic or not toxic. In more detail, a tweet is labeled as toxic if its \emph{toxicity} score is higher than or equal to 0.7, according to Google's Perspective API recommendation~\cite{jigsaw2018perspective}. 
We acknowledge that Google's Perspective API, as a toxicity detection tool, also has limitations~\cite{nogara2023toxic,teblunthuis2023misclassification, sap-etal-2022-annotators}, leading to the use of a stricter threshold, intending to reduce bias in the results. However, prior research has demonstrated its effectiveness in identifying various forms of toxic language in generated text as well as on social media data~\cite{ovalle2023m,kumarswamy2025causal,aleksandric2024users}. 

Consequently, around 2.4\% of the tweets in our dataset were considered as \textit{toxic}, while 56.2\% of the toxic tweets were posted by users other than root authors. 
The analysis focuses on toxic conversations only, as the bystander effect can only occur in cases where a toxicity attack exists within a conversation. 
Therefore, conversations that did not receive any toxic reply are discarded from the analysis, leaving the number of conversations at 10,455. In addition, we discarded 1,223 conversations that the root author initiated with the toxic reply (root tweet is toxic), as such conversations might not experience the same bystander effect as other conversations, which could possibly alter our results.

Also, note that some of the root authors appeared multiple times in our data, which could potentially introduce bias to our results since we use their account characteristics as control variables. Therefore, we randomly picked a single conversation from each user whose initial tweet repeatedly appeared in the dataset. \emph{Hence, the final dataset consists of 9,107 conversations consisting of 187,658 tweets posted by 118,609 unique users, where 5,115 conversations belong to \emph{dataset 1} and  the rest of 3,992 conversations originated from \emph{dataset 2}.}

\section{Methodology}
We use multivariate regression analysis to test our hypotheses.
Below, we explain our dependent, independent, and control variables in detail.

\textbf{Independent Variables:} The following independent variables are defined: (1)~\emph{users\_before\_toxic:} the number of unique users engaged in the conversation before the first toxic comment occurred. Examples of such tweets are colored in green in Figure~\ref{fig:twittergraph}. Note that even though some replies are at a higher level in the conversation tree, it does not necessarily mean that they occurred before the first toxic comment. For example, it is visible in Figure~\ref{fig:twittergraph} that the second layer of the tree contains both green and unlabeled nodes. Even though some of the unlabeled nodes are positioned higher in the tree structure, they are still posted later than the green nodes in lower levels of the tree.
Consequently, this variable was computed by chronologically ordering all tweets in the conversation from the oldest to the youngest and then calculating the number of unique users that posted tweets older than the first toxic tweet. 
Note that, on Twitter, with every new post, notifications are sent to users who have already participated in the discussion. 
For this variable, we could instead use the number of followers each victim has, but knowing that their accounts are public suggests that more people might be observing the conversation. Furthermore, followers might be observing the conversation; however, because they did not engage before the first toxic tweet, it is hard to claim they are interested in it. Therefore, we use the number of followers as a control variable in our regression models. (2)~\emph{reply\_toxicity:} the toxicity score of the first comment posted that is a direct reply to the first toxic comment in the conversation, e.g., circled in green in Figure~\ref{fig:twittergraph}. 

\textbf{Dependent Variables:} To test our hypotheses, we defined the following dependent variables:
(1)~\emph{unique\_users\_non\_toxic}: the total number of unique users who posted non-toxic comments after the first toxic reply in the conversation thread. We focus only on the single conversation thread that emerged from the first toxic reply, as circled in orange in Figure~\ref{fig:twittergraph}. We did not use the chronological ordering of all tweets in the conversation because we would capture users involved in other conversation threads, even though they were not related to the first toxic comment.
(2)~\emph{remaining\_toxicity}: the ratio of all toxic replies that occurred in the conversation thread after the first toxic reply and the total number of replies in the thread. Note that this calculation excludes the~\emph{reply\_toxicity}, and it is surrounded by the purple rectangle in Figure~\ref{fig:twittergraph}.

\textbf{Control Variables:} 
We controlled for several factors that could affect group behavior in Twitter conversations. We added controls for root authors' \emph{activity}, i.e., \emph{num\_friends}, \emph{num\_tweets}, and \emph{account\_age}, because more active users might have different audiences. For example, if a user posts many tweets, followers might engage less with their tweets and be less likely to defend them against toxicity attacks. 
We also controlled for the \emph{visibility}, i.e., \emph{num\_followers}, \emph{listed\_counts} and \emph{verified}. For example, users with \emph{verified} accounts or influencers with many followers might receive more help from others when they are under attack. 
Additionally, we controlled for profile characteristics, such as \emph{description\_length}, \emph{has\_URL}, and \emph{has\_location}. 
Users who provide less information on their profiles may receive less defense from other participants, as others may be less inclined to support anonymous users.
Finally, we used the \emph{width} and \emph{depth} of the conversation tree to control for differences in conversation structure. \emph{Depth} is the length from the root tweet to the conversation's deepest node, and \emph{width} represents the maximum number of tweets at any level in the conversation tree. 
Finally, control variables included in all the models were conversation topics. There is a possibility that the topics of conversations influence how toxic conversations unfold, as well as whether users encourage or stand up against toxicity. The list of topics is as follows: \emph{arts \& culture}, \emph{business \& entrepreneurs}, \emph{celebrity \& pop culture}, \emph{diaries \& daily life}, \emph{family}, \emph{fashion \& style}, \emph{film tv \& video}, \emph{fitness \& health}, \emph{food \& dining}, \emph{gaming}, \emph{learning \& educational}, \emph{music}, \emph{news \& social concern}, \emph{other hobbies}, \emph{relationships}, \emph{science \& technology}, \emph{sports}, \emph{travel \& adventure}, and \emph{youth \& student life}. Each topic with a coefficient greater than 0.5 was set to 1, and 0 otherwise (converting it to binary), thereby determining which topics each conversation is associated with. 

\begin{table*}[t!] \centering 
\caption{Descriptive statistics of variables used in the analysis} 
\label{table:1}
\resizebox{\textwidth}{!}{%
\begin{tabular}{l|l|llll|llll|llll}
\hline
& &\multicolumn{4}{|c|}{Merged Dataset} & \multicolumn{4}{|c|}{Dataset 1} & \multicolumn{4}{|c}{Dataset 2}\\
&Variable&Min&Median&Mean& Max&Min&Median&Mean& Max&Min&Median&Mean& Max\\
\hline
\hline
Dependent &remaining\_toxicity& 0 & 0 &  0.05 & 0.67 & 0 & 0 &  0.05 & 0.6 & 0 & 0 &  0.05 & 0.67 \\
Variables&unique\_users\_non\_toxic& 0 &  0 & 0.68 & 43 & 0 &  0 & 0.72 & 43 & 0 &  0 & 0.64 & 17 \\
\hline
Independent & users\_before\_toxic & 1 & 2 & 5.13 & 463 & 1 & 2 & 4.6 & 385 & 1 & 2 & 5.8 & 463 \\
Variables &reply\_toxicity& 0.001 & 0.18 & 0.25 & 0.97 & 0.001 & 0.18 & 0.25 & 0.97 & 0.005 & 0.19 & 0.26 & 0.96  \\
\hline
Control&num\_friends& 0  & 589 &  2006 & $\sim 1.5M$ & 0  & 560 &  1847.1 & $\sim 1.5M$& 0  & 642.5 &  2210.3 & $\sim 610.5K$\\
Variables &num\_tweets& 1 & $\sim 14K$  & $\sim 40K$ & $\sim 1.5M$ & 1 & $\sim 13K$  & $\sim 36K$ & $\sim 1.5M$ & 1 & $\sim 15.5K$  & $\sim 45K$ & $\sim 1M$ \\
&account\_age& 0  &  3 & 5.03 & 17 & 0  &  3 & 4.5 & 15 & 0  &  4 & 5.8 & 17 \\
&num\_followers& 0 & 1,233 & $\sim 97K$  & $\sim 55M$ & 0 & 1,019 & $\sim 58K$  & $\sim 54.4M$ & 0 & 1613 & $\sim 147K$  & $\sim 55M$ \\
&listed\_counts& 0  &  8 & 398.6  & $\sim 218K$ & 0  &  7 & 237.6 & $\sim 102.8K$ & 0  &  10 & 604.9  & $\sim 218K$ \\
&verified& 0  & 0 & 0.1  & 1 & 0  & 0 & 0.08  & 1 & 0  & 0 & 0.13  & 1 \\
&description\_length& 0  & 78 &  81.29  & 193 & 0  & 74 &  79.55  & 183 & 0  & 83 &  85.52  & 193 \\
&has\_URL& 0   &  0 & 0.5 &  1 & 0   &  0 & 0.5 &  1 & 0   &  0 & 0.5 &  1 \\
&has\_location& 0 &  1 &  0.76 &  1 & 0 &  1 &  0.76 &  1& 0 &  1 &  0.75 &  1\\
&width& 1 &   3 &  11.39 & 1,688 & 1 &   3 &  9.97 & 1,688 & 1 &   3 &  13.22 & 853 \\
&depth& 1 &   3 & 4.23 & 197 & 1 &   3 & 4.3 & 197 & 1 &   3 & 4.13 & 73 \\
\hline

\end{tabular}
}
\end{table*}

\section{Dataset Characterization} 
Our final dataset consists of 9,107 conversations with 187,658 tweets, posted by 118,609 unique users. Table~\ref{table:1} shows the descriptive statistics of the variables used in the statistical models. Note that we display the statistics of the final (merged) dataset, including the statistics on each dataset separately. \emph{Dataset 1} collected in 2021 includes 5,115 conversations, while \emph{dataset 2} collected in 2023 includes 3,992 conversations. The collection of two datasets allows us to gain insights into overall group dynamics and to check whether the trend may have changed over time.

The minimum number of tweets in conversations is 2, and the maximum is 1,689 tweets. The mean number of tweets included in these conversations is 20.6.
In more than half of conversations, the number of users engaged in the conversation before the first toxic reply (~\emph{users\_before\_toxic}) is 2, while the maximum number is 463. Also, the average of \emph{reply\_toxicity} is 0.25.
The mean number of \emph{unique\_users\_non\_toxic} is 0.68, indicating that in more than half of the conversations, there are users who posted non-toxic comments after the first toxic reply. The deepest conversation in our dataset, i.e., \emph{depth} is 197, while the maximum \emph{width} is 1,688. 

Interestingly, the maximum number of users posting non-toxic replies after a toxic reply in \emph{dataset 1} is 47, which is higher than that in \emph{dataset 2} (17). On the other hand, the maximum \emph{users\_before\_toxic} in \emph{dataset 2} is higher than that of \emph{dataset 1} {463 vs. 385}. Furthermore, the percentage of root authors in the two datasets who provided URLs and locations on their profiles does not differ significantly. However, Mann-Whitney tests indicate that all other root authors' characteristics differ significantly between the two datasets ($p < 0.05$). In addition, the differences in other dependent and independent variables between \emph{dataset 1} and \emph{dataset 2} are statistically significant ($p < 0.05$). 
In summary, the two conversation samples differed significantly, allowing us to examine whether our hypotheses hold across two time periods.
The top five topics in the dataset were \emph{celebrity \& pop culture}, \emph{film, tv \& video}, \emph{diaries \& daily life}, \emph{news \& social concern}, and \emph{sports}, which also matches the most prevalent topics in the two datasets separately.

\section{Results} 
To test hypotheses H1 and H2, we employed a Poisson regression model as the distribution of \emph{unique\_users\_non\_toxic} does not follow a normal distribution, and it is a count variable. We ran a linear regression model to test hypothesis H3, as \emph{remaining\_toxicity} follows a normal distribution. 
Additionally, we ran a Negative Binomial regression model that accounts for excess variance in count data, and our key findings remained consistent. 
 
\textbf{Higher levels of pre-toxic participation are associated with a lower number of users engaging in non-toxic responses following the toxic reply:}
The results obtained from a Poisson regression model are displayed in Table~\ref{results} (column H1) and reveal a statistically significant negative association between the \emph{unique\_users\_non\_toxic} and \emph{users\_before\_toxic} ($p < 0.001$). In other words, we find that the greater the number of users participating in the conversation before the first toxic comment is significantly correlated with the lower the number of users posting non-toxic replies after that comment, in support of \emph{H1}. 
Additionally, the results show that verified accounts are less likely to receive non-toxic comments from users after the first toxic reply than non-verified users, whereas the opposite is true for root authors who specified locations on their profiles. In other words, these results suggest that users are more likely to respond in a non-toxic way after toxicity when interacting with identifiable users, but are less likely to de-escalate when the root author is a verified account. It may be that highly visible accounts, such as verified users, are less likely to receive “standing up” responses to toxicity in their threads. Moreover, deeper conversations are more likely to have a higher number of users posting non-toxic replies. Interestingly, this implies that if a larger discussion develops, conversation participants tend to engage in a less toxic way.

\textbf{Greater toxicity level of the initial response to the first toxic reply is associated with less non-toxic engagement. }%
In our second hypothesis, we posited that the first reply immediately after the first toxic reply might play an important role in how the rest of the conversation develops. 
For this analysis, we discarded conversations that do not contain a direct reply to toxic tweets, leaving the dataset with 3,959 conversations, consisting of 75,926 tweets posted by 37,627 unique users.
\emph{Reply\_toxicity} was used as the independent variable in the Poisson regression model. Table~\ref{results} (column H2) shows that there is a negative statistically significant association between \emph{reply\_toxicity} and \emph{unique\_users\_non\_toxic} ($p < 0.001$). We find support for Hypothesis 2, with our results indicating that fewer users posting non-toxic replies is associated with greater toxicity of the first reply to the initial toxic comment.

\textbf{Toxic reply after first toxic reply is associated with more uncivil behavior: }
For our final hypothesis, H3, we examine how the toxicity of the first comment after the first toxic reply might determine the direction in which the conversation thread will emerge. In other words, if the first reply to a toxic reply is also toxic, the whole conversation thread might become more toxic. In this case, conversations that have less than two tweets after the first toxic reply are removed, leaving 2,230 conversations for analysis.
Furthermore, \emph{reply\_toxicity} was used as an independent variable in the linear regression model, while the dependent variable used was \emph{remaining\_toxicity}. The model results (Table~\ref{results} - column H3) indicate that there is a positive statistically significant association between \emph{reply\_toxicity} and \emph{remaining\_toxicity} ($p < 0.001$). This shows that there is a correlation between the higher toxicity of the immediate comment after the first toxic reply, 
and the greater toxicity levels in the remaining of the conversation thread. 
Thus, we find evidence in support of  \emph{H3}. 

To ensure that these findings remained consistent across samples, we ran models for both \emph{dataset 1} from 2021 and \emph{dataset 2} from 2023 individually. Once we determined that we found support for our hypotheses across datasets, the samples were combined and we presented findings from the merged dataset. 
This shows that our findings are not specific to a timeline and specific offline events and they can be generalized. 

\begin{table}[!htbp] \centering 
  \caption{Results of the regression analysis. Note that standard errors are presented in parentheses.} 
  \label{results} 
     \resizebox{\columnwidth}{!}{%

\begin{tabular}{@{\extracolsep{5pt}}lccc} 
\\[-1.8ex]\hline 
\hline \\[-1.8ex] 
 & \multicolumn{3}{c}{\textit{Dependent variable:}} \\ 
\cline{2-4} 
\\[-1.8ex] & \multicolumn{2}{c}{unique\_users\_non\_toxic} & remaining\_toxicity \\ 
\\[-1.8ex] & \textit{ H1: Poisson} & \textit{H2: Poisson} & \textit{H3 :OLS} \\
\hline \\[-1.8ex] 
 users\_before\_toxic & $-$0.02$^{***}$ (0.002) &  &  \\ 
  reply\_toxicity &  & $-$0.37$^{***}$ (0.058) & 0.05$^{***}$ (0.01) \\ 
  Followers & 0.0 (0.0) & $-$0.0 (0.0) & $-$0.0 (0.0) \\ 
  Friends & $-$0.00001$^{*}$ (0.0) & 0.00001$^{*}$ (0.0) & $-$0.0 (0.0) \\ 
  Num\_tweets & 0.0 (0.0) & 0.0 (0.0) & $-$0.0 (0.0) \\ 
  Listed\_count & 0.0 (0.00001) & 0.0 (0.00001) & 0.0 (0.0) \\ 
  VerifiedTrue & $-$0.45$^{***}$ (0.06) & 0.11 (0.06) & $-$0.01 (0.014) \\ 
  Age & $-$0.01 (0.003) & 0.004 (0.003) & $-$0.001 (0.001) \\ 
  UrlTrue & $-$0.04 (0.03) & $-$0.01 (0.03) & $-$0.01 (0.01) \\ 
  Description & 0.00003 (0.0003) & 0.001$^{*}$ (0.0003) & 0.0001 (0.0001) \\ 
  LocationTrue & 0.11$^{***}$ (0.03) & 0.02 (0.03) & $-$0.01 (0.01) \\ 
  Width & 0.0001 (0.001) & 0.0003 (0.0004) & $-$0.0002 (0.0001) \\ 
  Depth & 0.02$^{***}$ (0.001) & 0.01$^{***}$ (0.001) & 0.0001 (0.0003) \\ 
 \hline \\[-1.8ex] 
Observations & 9,107 & 3,959 & 2,230 \\ 
R$^{2}$ &  &  & 0.023 \\ 
\hline 
\hline \\[-1.8ex] 
\textit{Note:}  & \multicolumn{3}{r}{$^{*}$p$<$0.05; $^{**}$p$<$0.01; $^{***}$p$<$0.001} \\ 
\end{tabular} 
}
\end{table}

\section{Qualitative Analysis}

Even though our regression analysis shows a negative association between pre-toxic user participation and the number of users posting non-toxic reply post-toxicity, additional qualitative analysis could provide insights into how often standing up with the intent to resolve the conflict occurs in our dataset. 

\textbf{Labeling of the first toxic replies.} Firstly, we extracted a random sample of 200 conversations for the manual labeling, where at least one additional user was involved in the thread originating from the first toxic reply. The total number of such conversations is 3,959. This criterion ensures that standing up might be present, compared to conversations where no other users or replies are found after the first toxic reply. Then, two annotators labeled all the first toxic replies in these conversations with a binary label depending on whether it attacked an individual/group of people or not. The main idea is to filter out the conversations where the first toxic reply received a high toxicity score due to the usage of profanity words or attacks directed to non-human subjects, such as movies, anime, etc., as we are interested in positive standing up for human targets, as this scenario most closely resembles bullying compared to other scenarios mentioned. The two annotators selected for this task are computer science PhD students who are highly involved in social computing research, especially in content moderation and users' responses to online toxicity. The Cohen's Kappa~\cite{mchugh2012interrater} score for two sets of labels was 0.7, indicating substantial agreement. After the annotators met and resolved all conflicting labels, the total number of conversations in which the first toxic reply represented an attack against an individual or a group was 61.

\textbf{Labeling positive standing up.} The next step involved identifying conversations where positive standing up happens. Some examples of positive standing-ups are respectfully correcting misunderstandings, setting boundaries, disagreeing civilly, defending an individual or a group of people, etc.~\cite{allison2016cyber,biernesser2023middle,karna2011going,williford2012effects}. As shown in our regression analysis, toxicity begets more toxicity. Therefore, negative standing up, such as counterattacking, responding with sarcasm, threatening other users, and similar, is likely to escalate a conflict further rather than resolve it. Hence, in this analysis, we are interested in how often positive standing up with the intent to resolve the conflict occurs. Once again, two annotators manually labeled a thread emerging from the first toxic replies in these 61 conversations, assigning a binary label to each reply indicating whether it represented a positive standing up or not. The total number of replies labeled was 1,501, including the first toxic replies. The intercoder agreement Cohen's Kappa score equals 0.6, suggesting moderate agreement between annotators. This implies that this task is not easy for humans, potentially due to many replies in some threads, making it hard for annotators to understand the context of the conversation. Annotators met to resolve any conflicting labels. The total number of standing-up replies was 20 out of 1,440, belonging to 12 (19.7\%) conversations. These findings demonstrated that positive standing up with the intent to resolve the conflict rarely happens. 

\textbf{Representative examples of positive and negative standing up.} 
Below, we provide examples of positive and negative standing up that were identified by the coders in our dataset. Note that the reply that is in \textit{red} color signifies negative standing up, i.e., counterattacking and arguing their point of view by insulting others. Additionally, the reply in \textit{green} color demonstrated positive standing up, in this case, respectfully stating their point of view.

\emph{First toxic reply:} [MASK] You are not going to call the idiot an idiot He will gfight the Austin ISD all the way to the scummy TX Supreme Court and then your son won’t be safe Take a stand and call this pig out and his hypocrisy The moron is playing with Texans lives while getting the vaccine himself. 

\emph{First response:} \textcolor{green} {[MASK] [MASK] Getting the vaccine is a personal choice Same with wearing a mask One can choose to do so or not Respect the right to choose.}

\emph{Second response:}\textcolor{red} {[MASK] [MASK] No it’s not Are you stupid It impacts others so it is not You don’t live in a cave any more Is it your choice to shoot a gun on the street It’s not Get a clue Kids need to show multiple vaccinations before going to school camp etc.}

\section{Discussion}
The present study sheds light on the practical question of how norms of toxicity are established online. Online settings allow us to investigate how quickly such group norms are established and how closely one's actions conform to what is deemed appropriate or expected in the group~\cite{tajfel2010social}.Our study cannot measure users’ perceptions of the group nor their internalization of norms. However, we use a framework of social norms to gain insight into behavior, such as the number of participants prior to the first toxic reply and subsequent posting patterns, as proxies for the social context in which users are impacted. 
These behavioral patterns reveal how social dynamics within a single conversation can shape the nature of replies, highlighting the need for further research on the formation of descriptive social norms online.

We do not claim that the bystander effect is directly present in our dataset, as it is difficult to determine how many users actually observe a given incident (e.g., a toxic reply). However, when individuals see that no one intervenes in a toxic exchange, they may interpret this silence as signaling that such behavior is acceptable or that intervention may be further met with more antisocial behavior. This can create a feedback loop in which users look to others for cues, and the absence of response reinforces norms that tolerate or even legitimize toxic or antisocial behavior. This dynamic may suppress helping behavior, not only through diffusion of responsibility, but also by shaping perceptions of what constitutes appropriate online replies.

Despite these robust findings, the scope of this study was limited. Firstly, our sample was restricted to tweets in English. Secondly, our dataset does not capture the 'true' number of users who observe the conversation. Furthermore, what different users see on their feeds might be algorithmically selected to recommend the root post or relevant replies. Due to this, it is impossible to assume that a typical user will consume all the threads in the conversation or all the replies that occurred before the first toxic comment. Thus, it is hard to estimate the number of conversation observers. 
However, we believe \emph{users\_before\_toxic} can be a good proxy for the unique number of views because users who choose to participate are also more likely to observe how the rest of the conversation will unfold. Hence, they could potentially stand up for users who are attacked. 
In addition, the extent to which social media users counter toxicity can be influenced by factors such as the extremity of the views expressed~\cite{schieb2016governing}. 
Finally, we acknowledge that Google's Perspective API as a toxicity detection also contains certain limitations~\cite{nogara2023toxic,teblunthuis2023misclassification, sap-etal-2022-annotators}.
Despite these limitations, our results suggest that the perceptions of social group norms through increased conversational participants is associated with fewer users responding in a non-toxic manner to toxic replies. 
Furthermore, the similar trends observed across two datasets imply that this phenomenon was not chance-driven but rather reflects the true nature of online group social dynamics, which did not change over time.

Scholarship can build upon our work to investigate the decision-making process and the potential costs users face when considering whether to intervene or not to alter the tone of a Twitter conversation. For instance, users observing the conversation may struggle to empathize with targets of toxicity or hate speech because they lack insight into others’ perspectives or feelings. The literature showed that the bystanders' empathic concern shapes motivation to act and intervene~\cite{machackova2015brief}. Some recent scholarship has examined how technological designs may encourage bystander interventions on cyberbullying online on a large scale~\cite{taylor2019accountability}. Future work should examine interventions such as fostering bystanders’ role-taking, strengthening perspective-taking, and empathy for targets of incivility~\cite{davis2017self}. 

\section{Conclusion}

In this study, we assessed a random sample of ~\emph{Twitter Conversations} to understand how social norms and the toxicity level of the first response to a toxic reply can lead to the disinhibition of users' toxic replies. Motivated by theory of social norms and by the bystander effect theory, we find that increased conversational participants are associated with fewer Twitter users standing up to a toxic reply. We also highlight the importance of initial responses to toxic tweets within a conversation. Our results demonstrate that there is an association between posting a toxic reply immediately after an initial toxic comment and an increased likelihood of the remaining of the conversation being more toxic.
Understanding how users respond to uncivil or abusive content helps reveal how social norms shape behavior online.

\section{Acknowledgments}
This material is based upon work supported by the National Science Foundation under grant No. 2309318. Any opinions, findings, conclusions, or recommendations expressed in this material are those of the author(s) and do not necessarily reflect the views of the National Science Foundation. We would like to thank Sayak Saha Roy and Nazanin Salehabadi for initiating the research and helping us in the initial phase of data collection.

\bibliography{aaai25}
\section{Ethics Checklist}

\begin{enumerate}

\item For most authors...
\begin{enumerate}
    \item  Would answering this research question advance science without violating social contracts, such as violating privacy norms, perpetuating unfair profiling, exacerbating the socio-economic divide, or implying disrespect to societies or cultures?
    \textcolor{blue}{Yes}
  \item Do your main claims in the abstract and introduction accurately reflect the paper's contributions and scope?
     \textcolor{blue}{Yes}
   \item Do you clarify how the proposed methodological approach is appropriate for the claims made? 
     \textcolor{blue}{Yes, in the Hypotheses and Methodology sections.}
   \item Do you clarify what are possible artifacts in the data used, given population-specific distributions?
     \textcolor{blue}{Yes}
  \item Did you describe the limitations of your work?
     \textcolor{blue}{Yes, in the Discussion section.}
  \item Did you discuss any potential negative societal impacts of your work?
     \textcolor{green}{NA}
      \item Did you discuss any potential misuse of your work?
     \textcolor{green}{NA}
    \item Did you describe steps taken to prevent or mitigate potential negative outcomes of the research, such as data and model documentation, data anonymization, responsible release, access control, and the reproducibility of findings?
    \textcolor{blue}{No, as we are not making the dataset public.}
  \item Have you read the ethics review guidelines and ensured that your paper conforms to them?
    \textcolor{blue}{Yes}
\end{enumerate}

\item Additionally, if your study involves hypotheses testing...
\begin{enumerate}
  \item Did you clearly state the assumptions underlying all theoretical results?
    \textcolor{blue}{Yes, in the Hypotheses section.}
  \item Have you provided justifications for all theoretical results?
    \textcolor{blue}{Yes, in the Results section.}
  \item Did you discuss competing hypotheses or theories that might challenge or complement your theoretical results?
   \textcolor{blue}{Yes, in the Discussion section.}
  \item Have you considered alternative mechanisms or explanations that might account for the same outcomes observed in your study?
    \textcolor{blue}{Yes, in the Results and Discussion sections.}
  \item Did you address potential biases or limitations in your theoretical framework?
    \textcolor{blue}{Yes, in Hypotheses and Discussion sections.}
  \item Have you related your theoretical results to the existing literature in social science?
    \textcolor{blue}{Yes, in the Discussion section.}
  \item Did you discuss the implications of your theoretical results for policy, practice, or further research in the social science domain?
    \textcolor{blue}{Yes, in the Discussion section.}
\end{enumerate}

\item Additionally, if you are including theoretical proofs...
\begin{enumerate}
  \item Did you state the full set of assumptions of all theoretical results?
    \textcolor{green}{NA}
	\item Did you include complete proofs of all theoretical results?
    \textcolor{green}{NA}
\end{enumerate}

\item Additionally, if you ran machine learning experiments...
\begin{enumerate}
  \item Did you include the code, data, and instructions needed to reproduce the main experimental results (either in the supplemental material or as a URL)?
    \textcolor{green}{NA} 
  \item Did you specify all the training details (e.g., data splits, hyperparameters, how they were chosen)?
    \textcolor{green}{NA}
     \item Did you report error bars (e.g., with respect to the random seed after running experiments multiple times)?
    \textcolor{green}{NA}
	\item Did you include the total amount of compute and the type of resources used (e.g., type of GPUs, internal cluster, or cloud provider)?
     \textcolor{green}{NA}
     \item Do you justify how the proposed evaluation is sufficient and appropriate to the claims made? 
    \textcolor{green}{NA}
     \item Do you discuss what is ``the cost`` of misclassification and fault (in)tolerance?
   \textcolor{green}{NA}
  
\end{enumerate}

\item Additionally, if you are using existing assets (e.g., code, data, models) or curating/releasing new assets...
\begin{enumerate}
  \item If your work uses existing assets, did you cite the creators?
    \textcolor{blue}{Yes.}
  \item Did you mention the license of the assets?
    \textcolor{blue}{No, as the dataset is publically available.}
  \item Did you include any new assets in the supplemental material or as a URL?
    \textcolor{blue}{No}
  \item Did you discuss whether and how consent was obtained from people whose data you're using/curating?
    \textcolor{blue}{No, as we are using publically available Twitter data.}
  \item Did you discuss whether the data you are using/curating contains personally identifiable information or offensive content?
    \textcolor{blue}{Yes, we mentioned that the data contains toxic conversations.} 
\item If you are curating or releasing new datasets, did you discuss how you intend to make your datasets FAIR (see \citet{fair})?
\textcolor{green}{NA}
\item If you are curating or releasing new datasets, did you create a Datasheet for the Dataset 
\textcolor{green}{NA}
\end{enumerate}

\item Additionally, if you used crowdsourcing or conducted research with human subjects...
\begin{enumerate}
  \item Did you include the full text of instructions given to participants and screenshots?
    \textcolor{green}{NA}
  \item Did you describe any potential participant risks, with mentions of Institutional Review Board (IRB) approvals?
    \textcolor{green}{NA}
  \item Did you include the estimated hourly wage paid to participants and the total amount spent on participant compensation?
    \textcolor{green}{NA}
   \item Did you discuss how data is stored, shared, and deidentified?
   \textcolor{green}{NA}
\end{enumerate}

\end{enumerate}

\end{document}